\documentclass[sigconf]{acmart}

\AtBeginDocument{%
  }

\begin{document}

\title{Who Gets Seen in the Age of AI? Adoption Patterns of Large Language Models in Scholarly Writing and Citation Outcomes}

\author{Farhan Kamrul Khan}
\affiliation{%
  \institution{New York University Abu Dhabi}
  \city{Abu Dhabi}
  \country{UAE}
}

\author{Hazem Ibrahim}
\affiliation{%
  \institution{New York University Abu Dhabi}
  \city{Abu Dhabi}
  \country{UAE}
}

\author{Nouar Aldahoul}
\affiliation{%
  \institution{New York University Abu Dhabi}
  \city{Abu Dhabi}
  \country{UAE}
}

\author{Talal Rahwan}
\authornote{Corresponding author}
\affiliation{%
  \institution{New York University Abu Dhabi}
  \city{Abu Dhabi}
  \country{UAE}
}

\author{Yasir Zaki}
\authornotemark[1]
\affiliation{%
  \institution{New York University Abu Dhabi}
  \city{Abu Dhabi}
  \country{UAE}
}

\renewcommand{\shortauthors}{Khan et al.}


\begin{abstract}
The rapid adoption of generative AI tools is reshaping how scholars produce and communicate knowledge, raising questions about who benefits and who is left behind. We analyze over 230,000 Scopus-indexed computer science articles (2021–2025) to examine how AI-assisted writing alters scholarly visibility across regions. Using zero-shot detection of AI-likeness, we track stylistic changes in writing and link them to citation counts, journal placement, and global citation flows before and after ChatGPT. Our findings reveal uneven outcomes: authors in the Global East adopt AI tools more aggressively, yet Western authors gain more per unit of adoption due to pre-existing penalties for “humanlike” writing. Prestigious journals continue to privilege more human-sounding texts, creating tensions between visibility and gatekeeping. Network analyses show modest increases in Eastern visibility and tighter intra-regional clustering, but little structural integration overall. These results highlight how AI adoption reconfigures the labor of academic writing and reshapes opportunities for recognition.
\end{abstract}


\maketitle

\section{Introduction}

The release of ChatGPT on November 23, 2022 marked a turning point in how people engage in knowledge production, including the high-stakes domain of academic writing. While blind reliance on the tool carries risks \cite{lit-shopov, Athaluri2023}, in combination with human input, generative AI can enhance productivity \cite{shakked-productivity} and improve linguistic coherence \cite{Song-eflwriting}, particularly for non-English speaking scholars. In this sense, large language models act as linguistic equalizers, helping authors from the Global East or non-Anglophone regions overcome long-standing disadvantages tied to language fluency and stylistic expectations. By leveling this surface of the playing field, generative AI promises to expand who can participate and be recognized in global academia.

Yet this promise is complicated by evidence that AI may also reinforce inequality. Prestigious journals often privilege “human-like” writing, positioning AI-assisted texts as less authentic or rigorous. Moreover, while Global East authors adopt AI tools more aggressively, Western authors reap disproportionately higher returns from adoption, due to pre-existing citation penalties for human-like writing. These tensions suggest that generative AI reduces surface-level barriers without disrupting the deeper structural inequalities of scholarly prestige, visibility, and recognition \cite{Watermeyer2024, esl_David, esl_Yakhontova_2020}.

Our study investigates these dynamics through Computer Science, a field where Global East authorship has grown substantially and provides a meaningful site for comparative analysis \cite{AlShebli_beijing}. By examining writing styles, citation outcomes, and global citation networks before and after ChatGPT, we ask:

\noindent
\textbf{RQ1: \textit{How has the adoption of AI-assisted writing tools, such as ChatGPT, evolved across regions in academic research?}}

\noindent
To this end, we analyze over 150,000 articles to track the rise (or fall) of AI-assisted or AI-generated content in the Computer Science articles. That is, we measure the shift of academic writing towards a higher degree of AI-likeness across hemispheres (Global East vs Global West) and continents, and identify which ChatGPT releases have been the most influential driver in this shift.

\noindent
\textbf{RQ2: \textit{What is the relationship between AI usage in writing and academic visibility, particularly in terms of authorship citation metrics and publications in higher tier journals?}}

\noindent
We explore the causal relationship between AI-likeness and higher citation rates, and whether this trend has a regional variance. Specifically, we look at: a) whether AI-assisted authors generally receive more citations, b) whether AI-usage correlates to an increase in cross-regional citation flows (e.g. are Eastern authors getting cited more frequently by authors in the West, and vice versa?), and c) whether AI-usage increases the probability that one gets published in a higher-tier journal.

\noindent
\textbf{RQ3: \textit{How has the structure of the academic citation network changed since the introduction of ChatGPT, and what does it reveal about epistemic clustering or diffusion?}}

\noindent
In this exploratory analysis, we construct and look at citation networks within Computer Science research at a country level aggregate. We evaluate key network properties–including centrality, connectivity and conductance–both pre and post the release of ChatGPT in order to understand the broader influence dynamic, and observe trends towards marginalization or integration within the academic system.

\section{Related Work}
\subsection{AI Detection Tools}
Concerns over originality and unmonitored use of Generative AI tools have led to the proliferation of detection tools aiming to distinguish AI-generated text from human-authored ones. A core challenge identified across studies is the difficulty of creating robust, generalizable methods owing to the diverse and adaptive nature of generated content \cite{HansBinoculars, verma2024ghostbusterdetectingtextghostwritten, mao2024raidargenerativeaidetection}. 

Verma et al.~\cite{verma2024ghostbusterdetectingtextghostwritten} employ structured feature selection based on weaker LLMs' probabilities, which yields high accuracy within training domains while struggling with generalization across varying LLM architectures and content types. The model-specific training dependence limits its abilities to adapt to the evolving door of LLMs. To address this, Mao et al.~\cite{mao2024raidargenerativeaidetection} propose RAIDAR, a rewriting based detection tool relying on minimal feature dependence which exploits the intrinsic stability of AI generated texts, demonstrating a higher degree of invariance under LLM assisted rewriting compared to human generated text. The feature agnostic, rewriting-based design makes it robust against adversarial attempts at evasion, however the lower accuracy derived in some domains are likely stemming from some of their domain-specific datasets using singular-sentence rewrites instead of full-text, despite the paper claiming that shorter texts are harder to detect. 

In contrast, Hans et al.~\cite{HansBinoculars} demonstrate a different approach: contrasting the perplexity of a strong LLM model termed as the 'performer' (Falcons 7b-instruct) with the cross-perplexity of a weaker performer model (Falcon 7b), effectively capturing the relative 'humanness' of a text without explicit training on specific LLM outputs. The paper claims comparable accuracy to commercially available or research models (GPTZero, Ghostbuster), however, its the continuous metric of 'humanness' that is of our interest. Furthermore, the LLM models used are open source, and can be used without exorbitant API costs compared to the other two models, making it more suitable for our analysis.

\subsection{AI In Academia}
Corpus-level detection studies leave little doubt that Generative AI has moved from fringe curiosity to mainstream scholarly instrument. Working with one million arXiv abstracts (2018-2024), Geng \& Trotta estimate that the fraction of LLM-style abstracts is approximately 35\%” in computer science, using an adaptive word-frequency model calibrated on GPT-3.5 revisions~\cite{geng2024chatgpttransformingacademicswriting}. A convergent, distributional-quantification framework applied to 951 k papers by Liang et al. show a steady post-2022 rise in LLM-modified sentences, peaking at 17.5\% of CS abstracts and spreading, albeit more slowly, to physics and \textit{Nature} family journals~\cite{liang2024mappingincreasingusellms}. Extending the lens to pre-prints across \textit{arXiv}, \textit{bioRxiv} and \textit{medRxiv}, Cheng et al~\cite{Cheng2024Infiltrate} confirm “the widespread influence of AI-generated texts in scientific publications,” and link higher AI-text fractions to elevated citation counts. 

Beyond simple prevalence, scholars are beginning to test who benefits from LLM assistance. Using a 2.8 million article difference-in-differences design, Lin et al show that ChatGPT “significantly enhances lexical complexity in NNES-authored abstracts,” thereby narrowing a long-standing linguistic gap in scientific communication \cite{lin2025chatgptlinguisticequalizerquantifying}. The positive citation premium reported by Cheng et al suggests that stylistic upgrading may translate into greater scholarly visibility, though causal pathways remain contested~\cite{Cheng2024Infiltrate}. Overall, large-scale evidence now links Generative AI use to both measurable stylistic convergence and tentative gains in research impact, particularly for authors working outside the Anglophone core. 

Detection metrics tell only half the story; surveys illuminate how academics experience the Generative AI turn. A UK study of 284 scholars by Watermeyer et al.~\cite{Watermeyer2024} find that while many view Generative AI as a possible “source of positive disruption,” the tools “no more alleviate than extend the dysfunctions of neoliberal logic” and work intensification. In a different institutional setting, Al-Zahrani et al.\ report that 505 Saudi students exhibit high awareness and optimism about Generative AI’s research value but simultaneously argue for “the importance of adequate training, support, and guidance … [and] ethical considerations” \cite{Al-Zahrani2024_ImpactofGenAIonResearch}. These surveys demonstrate that users both wish to adopt Generative AI for higher productivity, while expecting clearer governance and adequate skill development.

\section{Methodology}

\subsection{Data Collection}

The dataset used in this project was collected through the Scopus Search and Abstract Retrieval APIs available via the Elsevier Developer Portal. In total, the dataset consists of 238,218 articles published between January 2021 and January 2025, all of which fall under the “Computer Science” category as identified using the ``comp'' subject filter. Each record includes associated bibliographic metadata, and, where available, the article abstract.

From this initial corpus, 155,177 articles were successfully matched to the Scimago Journal and Country Rank (SJR) dataset released in 2024. These matched records served as the basis for the controlled regression analyses, ensuring that journal-level characteristics could be incorporated into the modeling framework. For the exploratory network analysis, the entire dataset of 238,218 articles was utilized to maximize coverage of global citation flows. Regional classifications followed the continental mapping provided in the SJR dataset, with additional countries manually assigned when not present in the original mapping. To improve consistency across categories, the “Middle East/Africa” grouping—which in the original SJR dataset contained only Egypt—was expanded to encompass the broader set of Middle Eastern countries.

\subsection{AI-likeness Detection} \label{sec:binoculars}
As stated earlier, the model we use for Zero shot detection of AI generated text was Binoculars by Hans et. al (2024) \cite{HansBinoculars}. It provides a continuous 'Binoculars\_Score' metric that measures the intrinsic \textit{humanness} of an article, with the presupposition that human-like articles are more likely to be similar to weaker LLM models in terms of perplexity. The models we used are what's given in the default implementation, i.e., Falcon 7b as the observer and Falcon 7b-instruct as the performer. We've also used their binary ``accuracy\_prediction'' value (0 = human, 1 = AI), at their optimized threshold for F1.

\subsection{Adoption and Influence of Model Releases}
Given Binoculars score measures the ``humanness'' of a text in the original scale, we first trace out the average monthly binoculars score per regional unit across time on an inverted y-axis in Figure \ref{fig:gpt-evolution}, which effectively gives us an overall image of how the stylistic markers have changed over time in favor of being more/less AI like. Second, we run a regression with the following specification:

\begin{equation}
\begin{aligned}
B_i =\ & \alpha_0 + \alpha_1 G_i + \alpha_2 R_i + \alpha_3 (R_i \cdot G_i) + \boldsymbol{\gamma}' \mathbf{X}_i + \varepsilon_i
\end{aligned}
\label{eq:regional_style_shift}
\end{equation}

where we track whether the stylistic patterns are significantly different overall in comparison to a Pre-GPT baseline. Using standard regional dummies and interacting them with the GPT adoption indicator (i.e., $R_i \cdot G_i$), we track this for all regional units. $B_i$ denotes the binoculars score for article $i$, $G_i$ is a binary indicator for whether the article was published post-GPT, $R_i$ is a vector of regional dummy variables (e.g., hemisphere or continent), and $\mathbf{X}_i$ includes standard controls such as journal tier (SJR quartile) and author h-index.

Third, we measure how influential each model has been in driving this stylistic change on the overall dataset. We do this by introducing a categorical \texttt{gpt\_wave} variable for the 4 releases of focus, GPT 3.5, GPT 4, GPT 4o, and GPT 4o mini. The model specification is as follows:

\begin{equation}
\begin{aligned}
B_i =\ & \delta_0 + \sum_{k=1}^4 \delta_k D_{ik} + \boldsymbol{\theta}' \mathbf{X}_i + \eta_i
\end{aligned}
\label{eq:gpt_wave_effect}
\end{equation}

Here, $D_{ik}$ represents a set of dummy variables capturing the effect of each specific GPT model release $k$ (1 = GPT-3.5, 2 = GPT-4, etc.), with Pre-GPT articles serving as the reference category. The goal is to quantify the magnitude of stylistic change after each release, controlling for author- and journal-level characteristics through $\mathbf{X}_i$.

\subsection{Citation and Journal Tier Modeling, and inferring Causality}

\textbf{ZINB-Modeling for Citations}

Citation data, by design, is zero inflated and highly right skewed. That is, most articles will likely never get cited, and at the same time, if they do get cited, the number of citations are likely to be closer to zero. Second, the count data violates Poisson's assumption of the equality of mean and variance. Therefore, the design choice has to be a zero inflated binomial regression model, specified as follows:

\begin{equation}\label{eq:cit_zinb}
\begin{aligned}
\log \mathbb{E}[Y_i \mid Y_i > 0] =\ 
& \beta_0 + \beta_1 B_i + \beta_2 G_i + \beta_3 H_i \\
& +\ \beta_4 (B_i \cdot G_i) + \beta_5 (B_i \cdot H_i) \\
& +\ \beta_6 (G_i \cdot H_i) + \beta_7 (B_i \cdot G_i \cdot H_i) \\
& +\ \boldsymbol{\gamma}' \mathbf{Z}_i
\end{aligned}
\end{equation}

\noindent
Here, $Y_i$ is the number of citations received by article $i$. $B_i$ refers to the article's \texttt{binoculars\_score}, which measures how human-like the writing is (lower values indicate more AI assistance). This variable is also replaced with the binary predictor \texttt{accuracy\_prediction} in some models, which as stated in Section \ref{sec:binoculars} refers to a definitive prediction whether an article is AI written.
$G_i$ is a binary variable indicating whether the article was published after the release of ChatGPT.  
$H_i$ identifies whether the author is from the Global West (1) or Global East (0).  
$\mathbf{Z}_i$ includes other control variables such as the author's h-index, journal ranking (SJR quartile), and article exposure time.

In order to model the relationship of citation behavior with time, we construct a time-based exposure score that we derive from Wang et al.~\cite{wang_barabasi2013} model. The model specifies the probability of a paper $i$ being cited at time $t$:

\[
P_i(t) \propto h_i \cdot c_i(t) \cdot f(t)
\]

where $h_i$ is the fitness of the paper, $c_i(t)$ is the cumulative citations (preferential attachment), $f(t)$ is the aging function, modeled as a log-normal distribution.

And the log-normal aging function is defined as:

\[
f(t) = \frac{1}{\sqrt{2\pi} \sigma t} \exp\left( -\frac{(\ln t - \mu)^2}{2\sigma^2} \right)
\]

To quantify total exposure to citations up to age $T$, we integrate $f(t)$ from $0$ to $T$:

\[
\text{Exposure}(T) = \int_0^T f(t) \, dt
\]

Using a change of variables $z = \frac{\ln t - \mu}{\sigma}$, this becomes the cumulative distribution function (CDF) of a standard normal distribution:

\[
\text{Exposure}(T) = \Phi\left( \frac{\ln T - \mu}{\sigma} \right)
\]

This yields a scalar value between $0$ and $1$ representing the fraction of a paper's total citation opportunity it has had by age $T$.

In parallel to our citation‐count framework, we also examine whether “human‐likeness” predicts an article’s placement in journal tiers via an ordered‐logit specification.  Let \(Q_i\in\{1,2,3,4\}\) denote the SJR quartile of paper \(i\).  We model the cumulative logit as

\begin{equation}\label{eq:sjr_logit}
\begin{split}
\text{logit}\bigl(\Pr(Q_i \le k)\bigr)
= \tau_k
- \beta_1 B_i
- \beta_2 G_i
- \beta_3 H_i
- \beta_4 h_i \\
- \sum_{m=1}^{M} \delta_m\,\mathbb{1}\{\texttt{pub\_month}_i = m\},
\end{split}
\end{equation}

\noindent
for cutpoints \(\tau_1<\tau_2<\tau_3\).  Here \(B_i\), \(G_i\), \(H_i\), and \(h_i\) are as defined in Equation \ref{eq:cit_zinb}, and the \(\mathbb{1}\{\texttt{pub\_month}_i=m\}\) terms absorb seasonal journal cycles. A negative \(\beta_1\) therefore implies that, holding GPT era, author region, author prominence, and submission season fixed, more human‐like writing (\(\uparrow B_i\)) raises the odds of placement in a higher‐rank (lower‐numbered) quartile.  

\begin{figure}
    \centering
    \includegraphics[width=0.75\linewidth]{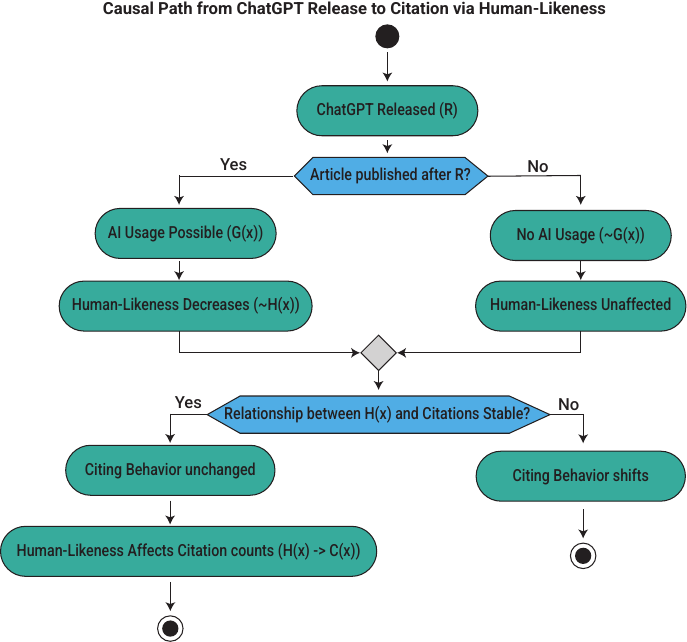}
    \caption{A Diagrammatic Depiction of the Path to Causality}
    \label{fig:causality-diagram}
    \Description{A Diagrammatic Depiction of the Path to Causality}
\end{figure}

Secondly, using the release of ChatGPT as a policy shock has its own endogeneity concerns when it comes to mediating citations. A \textit{post\_gpt} dummy will not only be predict citation behavior through binoculars scores, but also through its mediating relationship with the exposure control we modeled above. The argument for causality will not be made through instrumentation, but through the following logical path (see Figure~\ref{fig:causality-diagram}):

\noindent
\textbf{Premise 1: }All post-ChatGPT articles are necessarily younger than pre-ChatGPT ones.

\noindent
\textbf{Premise 2: }ChatGPT reduces human-likeness of Text.

\noindent
\textbf{Premise 3: }If ChatGPT is released, it can potentially influence all articles written afterwards–i.e., there is an aggregate effect of reduced human-likeness.

\noindent
\textbf{Premise 4: }Citing behavior with respect to human-likeness remains approximately stable before and after ChatGPT's release.

\noindent
\textbf{Conclusion:} Through its impact on human-likeness, the release of ChatGPT indirectly influences citation behavior.

\subsection{Network Metrics}

In our exploratory network analysis we employ five complementary measures to capture different facets of the global citation graph.  First, PageRank quantifies each country’s overall visibility by accounting for both the number and importance of incoming citation links.  Second, betweenness centrality identifies “broker” countries that lie on many shortest paths, highlighting key intermediaries in cross‐regional knowledge flows.  Third, conductance measures the proportion of citations that cross between predefined regions (e.g.\ East→West) versus remain within the same region, thus indexing the strength of regional silos.  Fourth, K‐core decomposition partitions the network into nested shells of increasingly dense connectivity, allowing us to track which countries move from the periphery into the structural core.  Finally, regional assortativity captures the tendency for countries to cite within their own region rather than cross‐regionally, providing a summary statistic for clustering by geography.  Together, these metrics allow us to trace how the release of ChatGPT may have reshaped both global visibility and local cohesion in academic citations.

\section{Results}
\subsection{RQ1: Investigating AI usage over time}

\begin{figure*}
    \centering
    \includegraphics[width=\linewidth]{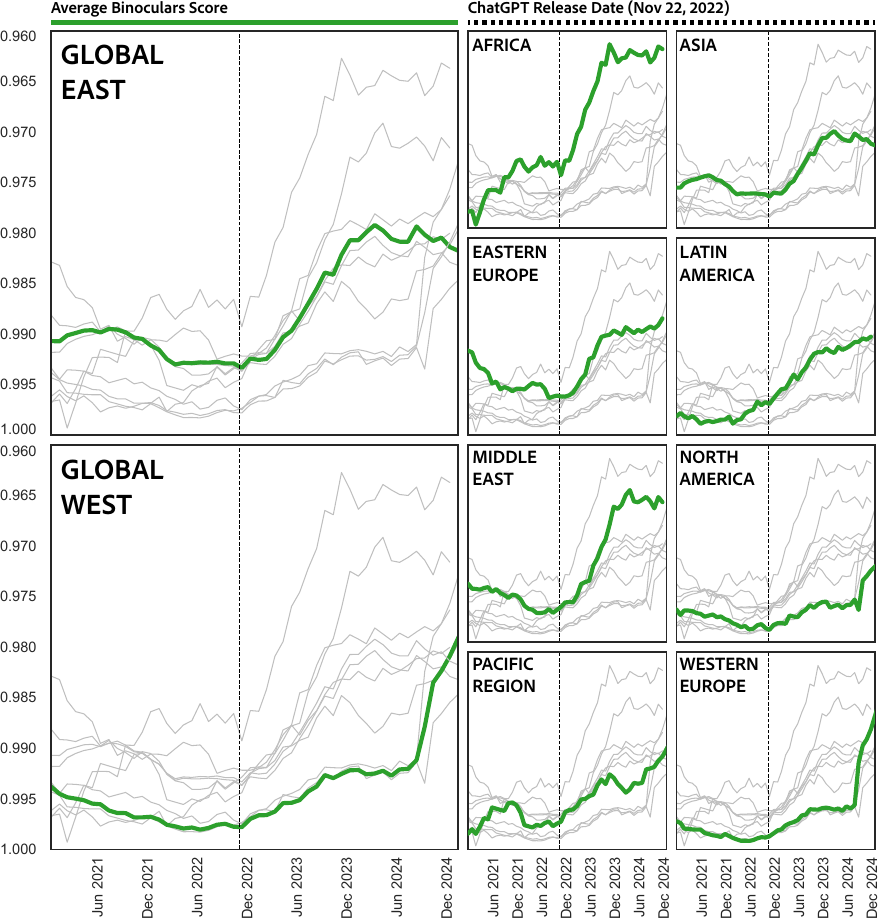}
    \caption{Evolution of Binoculars score over time on an inverted y-axis, separated by region. The trends suggest a global uptick in AI-ness in writing patterns, the change primarily driven by the Global East.}
    \label{fig:gpt-evolution}
    \Description{Evolution of Binoculars score over time on an inverted y-axis, separated by region. The trends suggest a global uptick in AI-ness in writing patterns, the change primarily driven by the Global East.}
\end{figure*}

The first objective of this study was to examine the overall uptick in AI usage across the years, using AI-ness as a proxy. Given Binoculars score indicates a measure of humanness in texts in the original axis configuration, we've inverted the y-axis in Figure \ref{fig:gpt-evolution} so that the visual uptick represents an increase in \textit{AI-ness} rather than \textit{humanness}. As can be seen, there is a clear difference in how the shift towards AI-ness manifests Post-GPT, with the Global East demonstrating a more pronounced shift. Most of the effects on the Global East seem to be mediated by Asia, while the Global West by North America, which is understandable given the volume of output from these regions.

To quantify this in the post-ChatGPT era, we employ equation \ref{eq:regional_style_shift} to regress Binoculars Score on a Post-GPT indicator, author hemisphere/region, and their interaction. This demonstrates a significant drop in human-likeness overall after ChatGPT's release (–0.00384, p < 0.001). The magnitude of this drop is significant, given that the Binoculars score hovers around a very small standard deviation (0.059), with a mean of 0.99. Authors in the Western hemisphere exhibit a small but significant higher baseline in human-likeness (+0.00174, p = 0.002). Critically, the PostGPT × West Hemisphere interaction (+0.00136, p = 0.068) is marginally significant, indicating that Western authors experienced a slightly smaller decline in humanness than Eastern authors. This pattern is visualized in the lower-left panel of Figure~\ref{fig:gpt-interaction}, where both hemispheres decline post-GPT, but the slope is steeper for the East; see Appendix Table~\ref{tab:hemisphere-model} for the full regression model. Thus, while AI-augmented writing has become more pervasive globally, Eastern authors appear to have adopted AI assistance more readily, altering their writing style to a greater extent.

We complement this analysis with a logistic model predicting the probability that a text is classified as AI-generated (1 = AI) by the same Post-GPT × Hemisphere structure. Post-ChatGPT texts, as a whole, are far more likely to be flagged as AI-like (+0.478 log-odds, p < 0.001). Neither the Western hemisphere main effect (0.0218, p = 0.691) nor its interaction (–0.0992, p = 0.139) reaches significance, implying that both hemispheres experienced comparable jumps in AI detectability. This is reflected in the lower-right panel of Figure~\ref{fig:gpt-interaction} and in Appendix Table~\ref{tab:logit-hemisphere}: both East and West rise from ~0.50 to ~0.60 probability post-GPT.

\begin{figure*}
    \centering
    \includegraphics[width=1\linewidth]{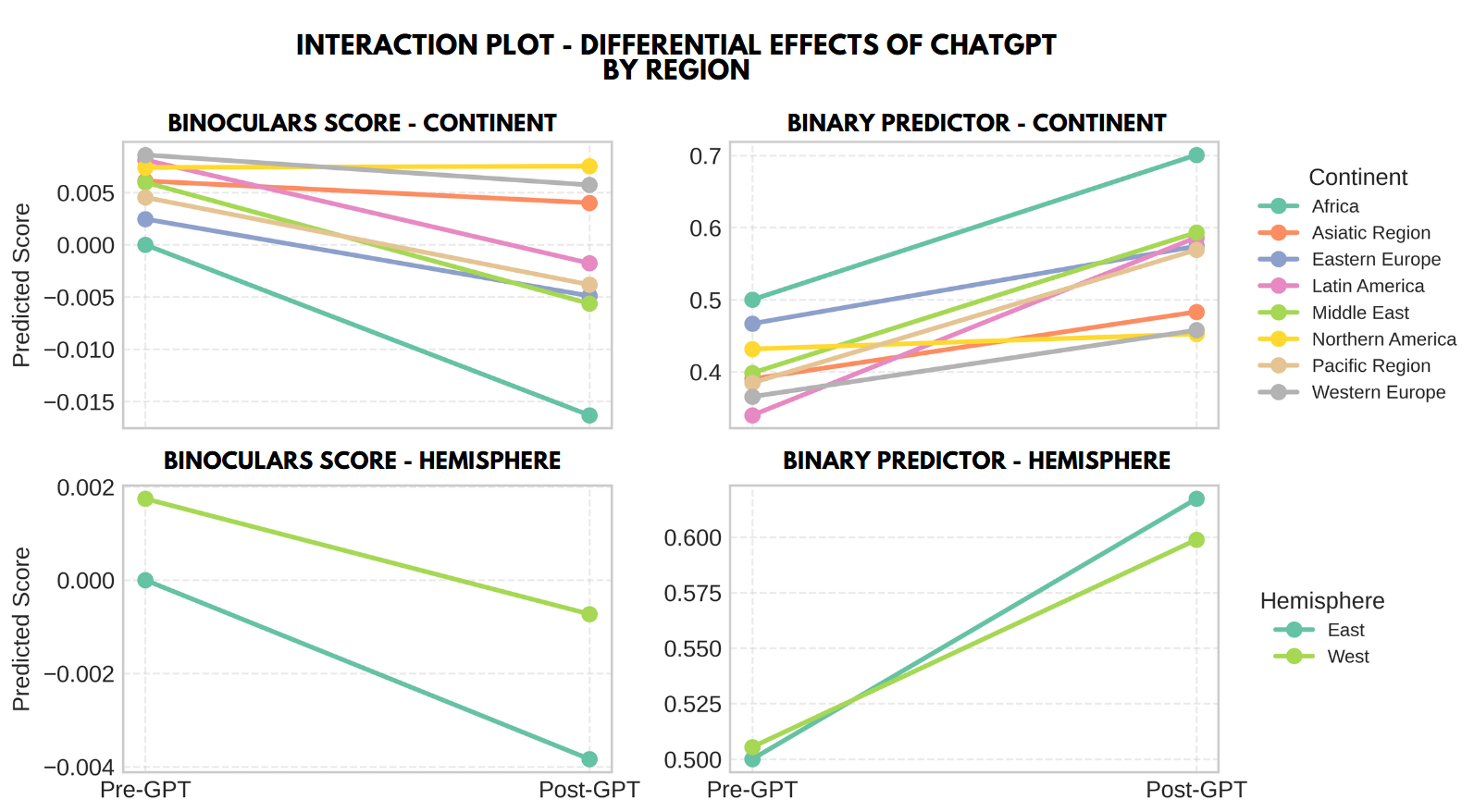}
    \caption{\textbf{Interaction Plot: ChatGPT Effects by Region.} Predicted “binoculars” centrality (left) and probability of a correct binary prediction (right) before (“Pre-GPT”) and after (“Post-GPT”) ChatGPT, stratified by continent (top) and hemisphere (bottom). Lines show region-specific marginal effects from interaction models. Note the pronounced decline in structural visibility (binoculars score) for Eastern‐hemisphere and African authors, alongside their larger gains in predictive accuracy compared to Western counterparts.}
    \label{fig:gpt-interaction}
    \Description{Predicted “binoculars” centrality (left) and probability of a correct binary prediction (right) before (“Pre-GPT”) and after (“Post-GPT”) ChatGPT, stratified by continent (top) and hemisphere (bottom). Lines show region-specific marginal effects from interaction models. Note the pronounced decline in structural visibility (binoculars score) for Eastern‐hemisphere and African authors, alongside their larger gains in predictive accuracy compared to Western counterparts.}
\end{figure*}

Extending to the 8 'continental' regions, the interacted OLS demonstrates that Asiatic Region ($\Delta$ = +0.0142, $p < 0.001$) and Northern America ($\Delta$ = +0.0164, $p < 0.001$) have the largest post-GPT declines in Binoculars Score relative to Africa.  Smaller but significant declines occur in Western Europe ($\Delta$ = +0.0134, $p < 0.001$) and Eastern Europe ($\Delta$ = +0.0090, $p = 0.005$); see Appendix Table~\ref{tab:continent-model}. Figure~\ref{fig:gpt-interaction} (upper-left) maps these continental slopes, highlighting regional heterogeneity in AI adoption speed. 

Similarly, the continent-interacted logit reveals that while Africa shows the greatest absolute jump in AI-detection likelihood (+0.851 log-odds, $p < 0.001$), the North American Post×NA interaction (–0.767, $p < 0.001$) substantially dampens its net effect, and Western Europe (–0.469, $p = 0.014$) and Asiatic Region (–0.472, $p = 0.008$) also exhibit smaller net increases. These patterns appear in the upper-right panel of Figure~\ref{fig:gpt-interaction}, indicating that regions differ not just in stylistic convergence but also in how machine detectors perceive that convergence; see Appendix Table~\ref{tab:logit-continent} for the full regression table. 

Next, we investigate the cumulative influence of each GPT model release on academic writing style, measured through the Binoculars Score (where higher = more human-like) and a binary classification of AI-generated text (where 1 = predicted as AI). The objective is to evaluate not just directional trends, but also which GPT release induced the most pronounced stylistic shift, and whether those shifts are statistically distinguishable.

The OLS model on Binoculars Score confirms that GPT-3.5’s impact is negligible and non-significant ($\beta = 0.00007$, $p = 0.94$). The GPT-4 period shows a modest and only marginally significant increase in humanness ($\beta = 0.0022$, $p = 0.067$), while more recent models—GPT-4o ($\beta = -0.0019$, $p = 0.046$) and GPT-4o-mini ($\beta = -0.0050$, $p < 0.001$)—produce statistically significant reductions in human-likeness; see Appendix Table~\ref{tab:study2-ols}. The point estimates suggest growing effects across releases; however, the confidence intervals for GPT-4 and GPT-4o do overlap, cautioning against strong claims of distinct magnitude without formal pairwise comparison tests. That said, the drop from GPT-4 to GPT-4o-mini is larger and statistically separable, indicating that GPT-4o-mini may mark a distinct threshold in stylistic AI-likeness. 

The logistic regression on AI classification likelihood mirrors these findings. Relative to Pre-GPT text, GPT-3.5 ($\beta = 0.085$, $p = 0.13$, n.s.), GPT-4 ($\beta = 0.170$, $p = 0.005$), GPT-4o ($\beta = 0.367$, $p < 0.001$), and GPT-4o-mini ($\beta = 0.603$, $p < 0.001$) are all increasingly likely to be flagged as AI-generated. Yet here too, confidence intervals between GPT-4 and 4o overlap, suggesting incremental increases rather than discrete jumps in influence. The effect of GPT-4o-mini, however, stands apart as both statistically and substantively the most influential. The effects are summarized in Figure \ref{fig:model-influence} and in Appendix Table~\ref{tab:study2-logit}.

In sum, each GPT model appears to have nudged writing further toward AI norms, but only GPT-4o-mini shows a clearly distinct leap in reducing human-likeness and increasing AI detectability. While the directional progression is clear, the differences between intermediate GPT versions are suggestive but not definitive without stronger statistical separation.

\begin{figure*}
    \centering
    \includegraphics[width=1\linewidth]{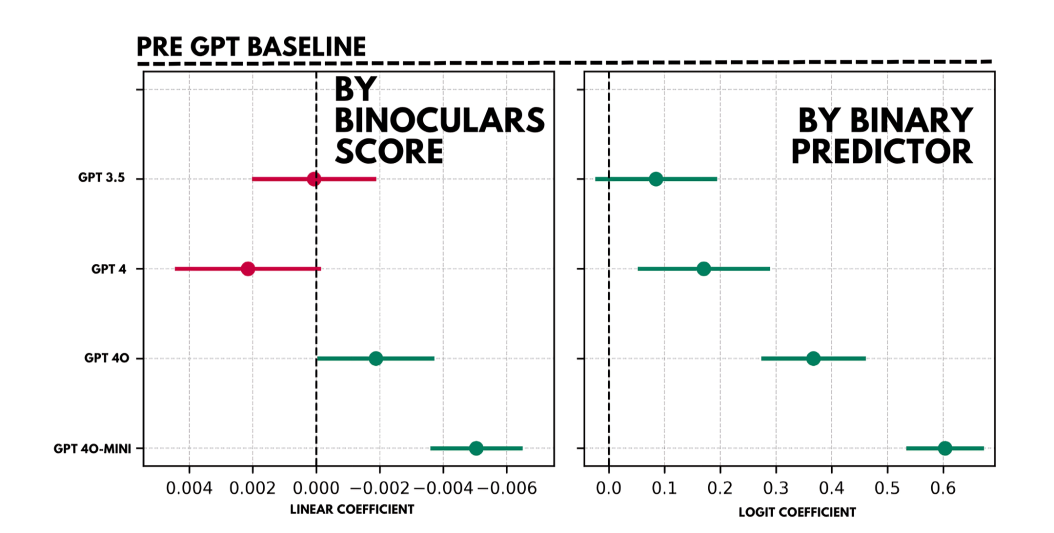}
    \caption{{GPT-Wave Effects Relative to Pre-GPT Baseline.} Point estimates and 95\% confidence intervals for the four GPT release dummies, plotted against a zero-coefficient Pre-GPT baseline (vertical dashed line). Left: OLS linear coefficients for Binoculars Score, illustrating progressive declines in human-likeness with each successive GPT model (sans GPT-4). Right: Logistic log-odds coefficients for the binary AI classification predictor, showing correspondingly increasing likelihood of being flagged as AI-generated post-GPT.}
    \Description{Point estimates and 95\% confidence intervals for the four GPT release dummies, plotted against a zero-coefficient Pre-GPT baseline (vertical dashed line). Left: OLS linear coefficients for Binoculars Score, illustrating progressive declines in human-likeness with each successive GPT model. Right: Logistic log-odds coefficients for the binary AI classification predictor, showing correspondingly increasing likelihood of being flagged as AI-generated post-GPT.}
    \label{fig:model-influence}
\end{figure*}

\subsection{RQ2: AI Usage and Citation Outcomes}\label{rq2}

Next, we look at whether there is a) a humanness penalty in citations, and b) if it has a differential effect on authors from the Global East in terms of citation visibility. To this end, we employ the zero inflated regression framework explained in Equation~\ref{eq:cit_zinb}. 

Across all groups and time periods, we find that greater 'humanness' is significantly associated with fewer citations ($\beta = -0.614$, p < 0.01). Western authors generally receive more citations overall ($\beta = 1.13$, p < 0.05), consistent with established intuition of a Western bias. But importantly, before GPT, Western authors faced harsher citation penalties for humanness ($\beta = -1.181$, p < 0.05) compared to their Eastern counterparts. The penalty however intensifies in the Post-GPT era ($\beta = -0.721$, p < 0.05), suggesting that the availability of ChatGPT may have raised the bar for 'natural' human writing in order to earn attention in the academic space. But the extra penalty for Western authors seems to have softened, although not significantly ($\beta$ = 1.150, p = 0.134). This indicates a shift towards convergence in stylistic penalties across both regions; see Appendix Table~\ref{tab:zinb-cited} for the full regression table.

In the Post-GPT era, therefore, Eastern authors have more to gain from using AI assistance in comparison to their Pre-GPT baseline. However, when viewed comparatively, Western authors derive greater benefit from AI-enhanced writing due to the pre-existing stylistic standards. The broader narrative of stylistic convergence is supported by the attenuation of humanness penalties Post-GPT for Western articles, the lack of significance in the three way interaction term makes in suggestive rather than definitive.

Shifting focus from total citation counts to cross-hemispheric citations, we ask whether more “human-like” (less AI-generated) writing affects visibility across regional boundaries. Using the same zero-inflated negative binomial framework, we test whether humanness imposes a cross-regional citation penalty, and whether such penalties vary across time and author location.

We find that greater humanness is again associated with a significant drop in citation visibility ($\beta = -1.79$, p < 0.001), indicating that the humanness penalty persists and is in fact more pronounced when visibility is measured across hemispheres. Unlike our earlier findings, however, the penalty does not significantly differ across time periods or author regions, as the three-way interaction between humanness, GPT exposure, and Western authorship is not significant ($\beta = -0.60$, p = 0.39); see Appendix Table~\ref{tab:zinb-citediff}.

These results suggest a shift: while stylistic convergence was visible in overall citations, cross-regional flows remain highly sensitive to AI-adaptive writing norms. In other words, "sounding human" may now be a greater liability when attempting to gain foreign attention, regardless of author region or GPT timing. The implications are stark, as AI conformity may be evolving into a global stylistic baseline for inter-regional visibility, rather than just a Western or post-GPT phenomenon.

As to the question of whether AI adoption impacts cross-regional citations, causality holds. Specifically, our causal argument demonstrated in Figure~\ref{fig:causality-diagram} rests on the key assumption that the citation behavior is stable across time in relation to binoculars score, that is, the interaction effect between Post-GPT dummy and Binoculars Score. For raw citations, the significance is weak ($\beta = -0.660$, p = 0.031), but for cross-region citations, it is largely insignificant (p = 0.299), implying that the shift in citations induced by changes in Binoculars score is attributable to the presence of GPT itself.

\subsection{Network Effects}\label{sec:network}

Lastly, we investigate structural shifts in the global computer science citation network around the release of ChatGPT, using a country-level aggregation of peer-reviewed publications. While we refrain from making strong causal claims, the analysis provides a descriptive portrait of evolving citation dynamics—highlighting changes in productivity, visibility, interregional flow, and positional hierarchy. Importantly, our dataset samples peer-reviewed literature indexed in Scopus, and while it is not exhaustive, country-level patterns are unlikely to diverge significantly from the full corpus.

We analyze the network using five structural measures: PageRank (visibility), betweenness centrality (brokerage), conductance (regional integration), K-core decomposition (core-periphery embedding), and regional assortativity (clustering by geography). These measures allow us to assess whether the rise of LLMs such as ChatGPT coincided with a measurable transformation in global citation topologies.

PageRank scores show a noticeable eastward tilt—China posts the sharpest gain (+0.0288), with Indonesia, Turkey, India, Uzbekistan, and Nigeria also rising, while the United States (-0.0069), United Kingdom (-0.0021), and Germany (-0.0021) slip slightly. Yet betweenness‐centrality shifts reveal a dual track: China (+0.0359) and India (+0.0284) are becoming major citation bridges, but the United States (+0.0265) and Germany (+0.0100) likewise strengthen their brokerage roles. The upshot is not a wholesale handover of influence but a widening of the core network: fast-growing Eastern actors join long-standing Western brokers, modestly rebalancing visibility while preserving a multipolar center of gravity. 

Post-GPT, regional assortativity more than doubled (0.028 → 0.074), indicating a stronger propensity for countries to cite within their own geographic blocs. Conductance patterns echo this: intra-Eastern flow climbed (0.848 → 0.857) and East-to-West leakage fell, while West-to-East citations ticked up but intra-Western ties weakened. The result is greater Eastern cohesion alongside slightly increased Western engagement, yet with higher mutual insulation between the two blocs. K-core analysis confirms the persistence of entrenched structures—only 3 of 25 peripheral countries moved to denser cores—signaling that the stylistic leveling afforded by LLMs has not, so far, disrupted the deep topology of global citation hierarchies. 

\begin{figure*}
    \centering
    \includegraphics[width=1\linewidth]{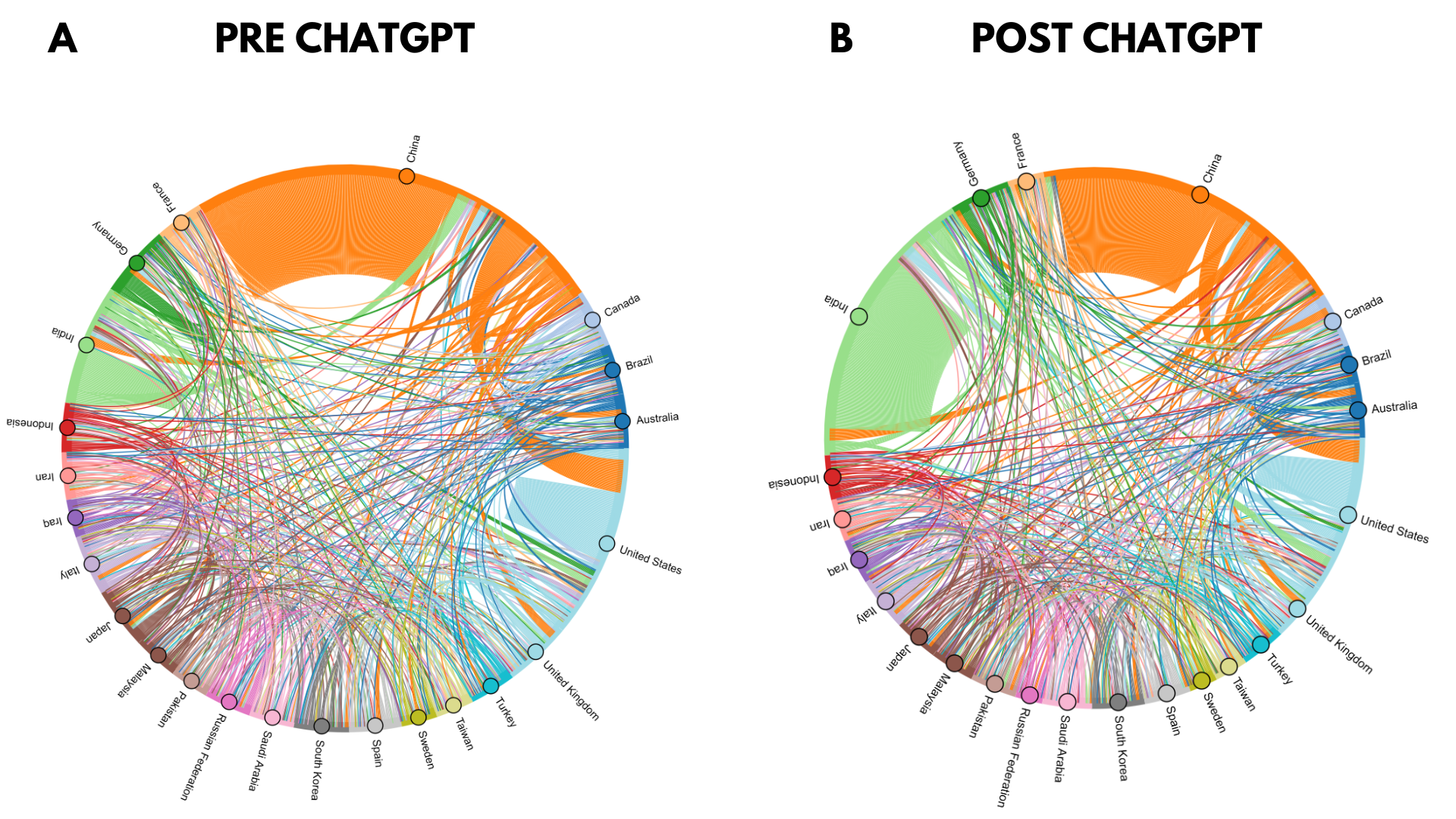}
    \caption{Chord diagrams depicting citation exchanges among the 20 most prolific countries in computer science research during (A) the Pre-ChatGPT period and (B) the Post-ChatGPT period. Each outer arc corresponds to a source country (arc length proportional to its total outgoing citations), and each ribbon represents the volume of citations flowing from the tail country to the head country (ribbon width $\propto$ citation count; ribbon color encodes the destination country). Note the pronounced enlargement of India’s arc and its incoming ribbons in (B), reflecting the growth of an intra-Eastern citation silo. Meanwhile, China’s dominant role persists, Eastern bloc countries exhibit tighter intra-regional linkages, and Western nations increasingly direct a greater share of their citations toward Eastern recipients, indicating a modest rebalancing of global visibility without displacing core actors.}
    \label{fig:chord}
    \Description{Chord diagrams depicting citation exchanges among the 20 most prolific countries in computer science research during (A) the Pre-ChatGPT period and (B) the Post-ChatGPT period. Each outer arc corresponds to a source country (arc length proportional to its total outgoing citations), and each ribbon represents the volume of citations flowing from the tail country to the head country (ribbon width $\propto$ citation count; ribbon color encodes the destination country). Note the pronounced enlargement of India’s arc and its incoming ribbons in (B), reflecting the growth of an intra-Eastern citation silo. Meanwhile, China’s dominant role persists, Eastern bloc countries exhibit tighter intra-regional linkages, and Western nations increasingly direct a greater share of their citations toward Eastern recipients, indicating a modest rebalancing of global visibility without displacing core actors.}
\end{figure*}

\section{Discussion}
\subsection{Implications}
This research builds upon a growing body of empirical work tracking the rising prevalence of Large Language Models (LLMs) in academic writing, particularly within the field of computer science. Previous studies have demonstrated corpus-level increases in LLM generated text, particularly with a notable concentration in Computer Science \cite{geng2024chatgpttransformingacademicswriting, geng2025humanllmcoevolutionevidenceacademic, liang2024mappingincreasingusellms}. However, most of these studies focus on detection or prevalence, rather than measuring the epistemic and structural implications of such a shift. Our work complements and extends these efforts by using a zero-shot AI detection framework to not only track changes in writing style, but also link those stylistic patterns to citation visibility, journal placement, and contextualize this within broader citation network dynamics. 

Unlike prior studies that sample across heterogeneous corpora (arXiv, bioRxiv or Nature for example), this project focuses exclusively on Scopus-indexed journals within computer science. This provides a curated, peer-reviewed dataset that captures institutional and editorial filtering more directly. This narrower focus allows us to interrogate the relationship between stylistic preferences and publication outcomes within formal academic gate-keeping environments. It also aids in isolating the effects of AI assistance within a domain that is both highly productive and has a more diverse geo-spatial representation compared to other disciplines. In doing so, we demonstrate that post-ChatGPT stylistic convergence correlates with measurable advantages in visibility, particularly for authors in the Global East, strengthening the hypothesis that LLMs are structurally reconfiguring who gets heard.

A key tension that emerges from both our data and the literature is in the divergent preferences of journal editors and researchers in regards to AI-assisted writing. Survey based studies demonstrate that researchers from underrepresented regions in particular hold an optimistic perspective of Generative AI tools as a means of boosting productivity and overcoming linguistic barriers. The same researchers however also express concerns over a lack of ethical guidance, training infrastructure, and editorial transparency in the use of such tools. What we do in our research is provide an empirical corollary to these concerns. The fact that the editorial structure appears more conservative, favoring traditionally "human-like" writing or perceiving AI-likeness as a sign of diminished intellectual rigor or authenticity. This is concerning, because our research demonstrates that even Pre-GPT, the baseline "human-ness" of Western authors are higher ($\beta$ = 0.0017, $p \leq 0.001$), corroborating the narrative of a specific stylistic preference. Fluency optimized for discoverability may not align with the stylistic expectations of elite gate-keeping institutions, or there may be internalized biases at play (see Appendix Table~\ref{tab:hemisphere-model}).

In the absence of transparent editorial guidelines, researchers may find themselves in a precarious position of being pressured to adopt stylistic conventions that improve visibility, yet get penalized for doing so when seeking out formal prestige. Furthermore, as Watermeyer et al. (2024) demonstrates, the rising pressure of productivity in academia is further exacerbated by the GPT induced productivity boost, and that in turn again makes it difficult for authors to not use Generative AI tools to accelerate productivity~\cite{Watermeyer2024}. Reconciling this split requires clearer signaling from journals regarding acceptable bounds of AI usage.

\subsection{Limitations}
Section \ref{sec:network} is largely observational, and the shifts observed are not necessarily linked to the release of ChatGPT but broader patterns. A lot of the changes are driven by the fact that Beijing has consistently been at the core of AI research over the years, and the center of mass of the AI research network had been seeing an Eastward shift in the last three decades \cite{AlShebli_beijing}. This is simply a part of a larger trend, although India's sudden growth in internal research productivity is something that is remarkable, and could be a signal of an emerging regional momentum that was not necessarily seen in prior research.

Next, the identification of a causal chain "ChatGPT $\rightarrow$ human-likeness $\rightarrow$ citations" hinges upon treating the ChatGPT release dummy as a quasi-instrument for writing style. First, the \textit{instrument} is not clean, and requires the assumption that citation behavior on humanness remains stable before and after the release of ChatGPT. Although we demonstrated that explicitly on a singular point, a leads-and-lags estimation would be able to pick up changes in behavior at other points in time–and the weak significance seen in our analysis of RQ1 suggests that this is worth looking into. Second, the research neither provides a population level estimate, nor a clean random sample level estimate. This is because of the pagination limits intrinsically restrict the number of articles we can extract at the most granular level (5000 articles for every month of data, maximum). Furthermore, the \textit{relevance} metric may introduce systemic biases into the kind of representation we have within our sample. Thirdly, the estimates don't account for when a citing article cites another, i.e., change in productivity levels and citation behavior might have influenced citation behavior in the pre-treatment group, meaning that the estimates might be at the lower bound. And lastly, the majority of detection tools place a baseline penalty on Global East authorship, i.e., work from Eastern authors are more likely to be predicted to have higher Generative AI aided content \cite{verma2024ghostbusterdetectingtextghostwritten, mao2024raidargenerativeaidetection}, however, our quasi-instrumental approach considers the baseline shift.


\bibliographystyle{ACM-Reference-Format}
\bibliography{sample-base}

\appendix

\begin{table}[!ht]
\centering
\scriptsize
\renewcommand{\arraystretch}{1.2}
\begin{tabular}{lcc}
\toprule
\textbf{Variable} & \textbf{Estimate (SE)} & \textbf{Signif.} \\
\midrule
Post-GPT & $-0.0038\ (0.0004)$ & $^{***}$ \\
Hemisphere: West & $0.0017\ (0.0005)$ & $^{***}$ \\
Post-GPT × West & $0.0014\ (0.0007)$ & $^{*}$ \\
SJR Q2 & $-0.0013\ (0.0005)$ & $^{**}$ \\
SJR Q3 & $-0.0024\ (0.0007)$ & $^{***}$ \\
SJR Q4 & $-0.0047\ (0.0007)$ & $^{***}$ \\
h-index & $4.3 \times 10^{-6}\ (2.5 \times 10^{-6})$ &  \\
\bottomrule
\end{tabular}
\begin{flushleft}
\footnotesize{$N=97,995$, $R^2=0.0031$, Root MSE = 0.055. $^{***}p<0.001$, $^{**}p<0.01$, $^{*}p<0.1$}
\end{flushleft}
\caption{Regression results predicting \textit{binoculars\_score} (Hemisphere Model). Robust standard errors in parentheses.}
\vspace{-1em}
\label{tab:hemisphere-model}
\end{table}

\begin{table}[!ht]
\centering
\scriptsize
\renewcommand{\arraystretch}{1.2}
\begin{tabular}{lcc}
\toprule
\textbf{Variable} & \textbf{Estimate (SE)} & \textbf{Signif.} \\
\midrule
Post-GPT & $0.478\ (0.038)$ & $^{***}$ \\
Hemisphere: West & $0.022\ (0.055)$ &  \\
Post-GPT × West & $-0.099\ (0.067)$ &  \\
SJR Q2 & $0.282\ (0.047)$ & $^{***}$ \\
SJR Q3 & $0.489\ (0.055)$ & $^{***}$ \\
SJR Q4 & $0.693\ (0.052)$ & $^{***}$ \\
h-index & $0.00003\ (0.00025)$ &  \\
\bottomrule
\end{tabular}
\begin{flushleft}
\footnotesize{$N = 97{,}995$, Pseudo-$R^2 = 0.0148$, Log Likelihood = $-19198.17$. $^{***}p<0.001$, $^{**}p<0.01$, $^{*}p<0.1$}
\end{flushleft}
\caption{Logistic regression predicting \textit{accuracy\_prediction} (Hemisphere Model). Robust standard errors in parentheses.}
\vspace{-1em}
\label{tab:logit-hemisphere}
\end{table}

\begin{table}[!hb]
\centering
\scriptsize
\renewcommand{\arraystretch}{1.2}
\begin{tabular}{lcc}
\toprule
\textbf{Variable} & \textbf{Estimate (SE)} & \textbf{Signif.} \\
\midrule
Post-GPT & $-0.0163\ (0.0029)$ & $^{***}$ \\
Africa/MidEast & $0.0148\ (0.0038)$ & $^{***}$ \\
Asia & $0.0061\ (0.0021)$ & $^{***}$ \\
East Europe & $0.0025\ (0.0025)$ &  \\
Latin America & $0.0081\ (0.0026)$ & $^{***}$ \\
Middle East & $0.0051\ (0.0023)$ & $^{**}$ \\
North America & $0.0074\ (0.0022)$ & $^{***}$ \\
Pacific & $0.0045\ (0.0028)$ &  \\
West Europe & $0.0086\ (0.0022)$ & $^{***}$ \\
Post × Africa/MidEast & $0.0011\ (0.0060)$ &  \\
Post × Asia & $0.0142\ (0.0030)$ & $^{***}$ \\
Post × East Europe & $0.0090\ (0.0035)$ & $^{**}$ \\
Post × Latin Am. & $0.0064\ (0.0037)$ & $^{*}$ \\
Post × Middle East & $0.0052\ (0.0033)$ &  \\
Post × N. America & $0.0164\ (0.0031)$ & $^{***}$ \\
Post × Pacific & $0.0080\ (0.0040)$ & $^{**}$ \\
Post × W. Europe & $0.0134\ (0.0030)$ & $^{***}$ \\
SJR Q2 & $-0.0005\ (0.0005)$ &  \\
SJR Q3 & $-0.0018\ (0.0007)$ & $^{***}$ \\
SJR Q4 & $-0.0048\ (0.0007)$ & $^{***}$ \\
h-index & $3.3 \times 10^{-6}\ (2.5 \times 10^{-6})$ &  \\
\bottomrule
\end{tabular}
\begin{flushleft}
\footnotesize{$N=97,995$, $R^2=0.0031$, Root MSE = 0.055. $^{***}p<0.001$, $^{**}p<0.01$, $^{*}p<0.1$}
\end{flushleft}
\caption{Regression results predicting \textit{binoculars\_score} (Continent Model). Robust standard errors in parentheses.}
\vspace{-1em}
\label{tab:continent-model}
\end{table}

\begin{table}[!hb]
\centering
\scriptsize
\renewcommand{\arraystretch}{1.2}
\begin{tabular}{lcc}
\toprule
\textbf{Variable} & \textbf{Estimate (SE)} & \textbf{Signif.} \\
\midrule
Post-GPT & $0.851\ (0.172)$ & $^{***}$ \\
Asia & $-0.447\ (0.153)$ & $^{**}$ \\
East Europe & $-0.133\ (0.186)$ &  \\
Latin America & $-0.665\ (0.228)$ & $^{**}$ \\
Middle East & $-0.412\ (0.175)$ & $^{**}$ \\
North America & $-0.275\ (0.165)$ & $^{*}$ \\
Pacific & $-0.470\ (0.264)$ & $^{*}$ \\
West Europe & $-0.551\ (0.164)$ & $^{***}$ \\
Post × Asia & $-0.472\ (0.178)$ & $^{***}$ \\
Post × East Europe & $-0.419\ (0.220)$ & $^{*}$ \\
Post × Latin Am. & $0.162\ (0.264)$ &  \\
Post × Middle East & $-0.062\ (0.206)$ &  \\
Post × N. America & $-0.767\ (0.195)$ & $^{***}$ \\
Post × Pacific & $-0.103\ (0.311)$ &  \\
Post × W. Europe & $-0.469\ (0.192)$ & $^{**}$ \\
SJR Q2 & $0.167\ (0.048)$ & $^{***}$ \\
SJR Q3 & $0.437\ (0.056)$ & $^{***}$ \\
SJR Q4 & $0.676\ (0.053)$ & $^{***}$ \\
h-index & $0.00005\ (0.00025)$ &  \\
\bottomrule
\end{tabular}
\begin{flushleft}
\footnotesize{$N = 97{,}162$, Pseudo-$R^2 = 0.0194$, Log Likelihood = $-18708.30$. $^{***}p<0.001$, $^{**}p<0.01$, $^{*}p<0.1$}
\end{flushleft}
\caption{Logistic regression predicting \textit{accuracy\_prediction} (Continent Model). Robust standard errors in parentheses.}
\vspace{-1em}
\label{tab:logit-continent}
\end{table}

\clearpage

\begin{table}[!ht]
\centering
\scriptsize
\renewcommand{\arraystretch}{1.2}
\begin{tabular}{lcc}
\toprule
\textbf{Variable} & \textbf{Estimate (SE)} & \textbf{Signif.} \\
\midrule
\multicolumn{3}{l}{\textit{GPT Wave}} \\
Post-GPT3.5 & $0.00007\ (0.00100)$ &  \\
Post-GPT4 & $0.00215\ (0.00117)$ & $^{*}$ \\
Post-GPT4o & $-0.00188\ (0.00094)$ & $^{**}$ \\
Post-GPT4o+ & $-0.00504\ (0.00074)$ & $^{***}$ \\
\addlinespace
\multicolumn{3}{l}{\textit{Global East}} \\
Global East & $-0.00196\ (0.00057)$ & $^{***}$ \\
\addlinespace
\multicolumn{3}{l}{\textit{Interaction: GPT Wave × Global East}} \\
GPT3.5 × East & $0.00051\ (0.00123)$ &  \\
GPT4 × East & $-0.00140\ (0.00145)$ &  \\
GPT4o × East & $-0.00193\ (0.00119)$ &  \\
GPT4o+ × East & $-0.00054\ (0.00091)$ &  \\
\addlinespace
\multicolumn{3}{l}{\textit{Controls}} \\
SJR Q2 & $-0.00123\ (0.00052)$ & $^{**}$ \\
SJR Q3 & $-0.00224\ (0.00066)$ & $^{***}$ \\
SJR Q4 & $-0.00464\ (0.00065)$ & $^{***}$ \\
h-index & $5.0 \times 10^{-6}\ (2.5 \times 10^{-6})$ & $^{**}$ \\
\bottomrule
\end{tabular}
\begin{flushleft}
\footnotesize{$N = 97{,}995$, $R^2 = 0.0043$, Root MSE = 0.055. $^{***}p<0.001$, $^{**}p<0.01$, $^{*}p<0.1$}
\end{flushleft}
\caption{OLS regression predicting \textit{binoculars\_score} across GPT wave and region (Global East). Robust standard errors in parentheses.}
\vspace{-1em}
\label{tab:study2-ols}
\end{table}

\begin{table}[!ht]
\centering
\scriptsize
\renewcommand{\arraystretch}{1.2}
\begin{tabular}{lcc}
\toprule
\textbf{Variable} & \textbf{Estimate (SE)} & \textbf{Signif.} \\
\midrule
\multicolumn{3}{l}{\textit{GPT Wave}} \\
Post-GPT3.5 & $0.085\ (0.056)$ &  \\
Post-GPT4 & $0.170\ (0.061)$ & $^{**}$ \\
Post-GPT4o & $0.367\ (0.048)$ & $^{***}$ \\
Post-GPT4o+ & $0.603\ (0.036)$ & $^{***}$ \\
\addlinespace
\multicolumn{3}{l}{\textit{Global East}} \\
Global East & $0.033\ (0.033)$ &  \\
\addlinespace
\multicolumn{3}{l}{\textit{Controls}} \\
SJR Q2 & $0.277\ (0.047)$ & $^{***}$ \\
SJR Q3 & $0.476\ (0.055)$ & $^{***}$ \\
SJR Q4 & $0.687\ (0.052)$ & $^{***}$ \\
h-index & $-0.00003\ (0.00025)$ &  \\
\bottomrule
\end{tabular}
\begin{flushleft}
\footnotesize{$N = 97{,}995$, Pseudo-$R^2 = 0.0175$, Log Likelihood = $-19145.95$. $^{***}p<0.001$, $^{**}p<0.01$, $^{*}p<0.1$}
\end{flushleft}
\caption{Logistic regression predicting \textit{accuracy\_prediction} across GPT wave and region (Global East). Robust standard errors in parentheses.}
\vspace{-1em}
\label{tab:study2-logit}
\end{table}

\begin{table}[hb]
\centering
\scriptsize
\renewcommand{\arraystretch}{1.2}
\begin{tabular}{lcc}
\toprule
\textbf{Variable} & \textbf{Estimate (SE)} & \textbf{Signif.} \\
\midrule
\multicolumn{3}{l}{\textit{Main Model}} \\
Binoculars Score & $-0.614\ (0.211)$ & $^{***}$ \\
Post-GPT & $0.857\ (0.310)$ & $^{***}$ \\
Post-GPT × Binoculars Score & $-0.721\ (0.308)$ & $^{**}$ \\
West & $1.129\ (0.535)$ & $^{**}$ \\
West × Binoculars Score & $-1.181\ (0.534)$ & $^{**}$ \\
Post-GPT × West & $-1.185\ (0.763)$ &  \\
Post-GPT × West × Binoculars & $1.150\ (0.767)$ &  \\
h-index & $0.0020\ (0.0001)$ & $^{***}$ \\
SJR Q2 & $-0.718\ (0.022)$ & $^{***}$ \\
SJR Q3 & $-1.352\ (0.028)$ & $^{***}$ \\
SJR Q4 & $-2.079\ (0.028)$ & $^{***}$ \\
Exposure Score & $2.178\ (0.057)$ & $^{***}$ \\
\midrule
\multicolumn{3}{l}{\textit{Inflation Model}} \\
Binoculars Score & $0.026\ (0.451)$ &  \\
h-index & $-0.0041\ (0.0003)$ & $^{***}$ \\
Exposure Score & $-92.623\ (4.939)$ & $^{***}$ \\
SJR Q2 & $0.210\ (0.073)$ & $^{***}$ \\
SJR Q3 & $0.772\ (0.087)$ & $^{***}$ \\
SJR Q4 & $1.856\ (0.124)$ & $^{***}$ \\
\bottomrule
\end{tabular}
\caption{Zero-inflated negative binomial model predicting overall citation count.}
\vspace{-1em}
\begin{flushleft}
\footnotesize{$N = 97{,}995$, Log Likelihood = $-251{,}658$, Wald $\chi^2 = 15{,}792$, $p<0.001$. $^{***}p<0.001$, $^{**}p<0.01$, $^{*}p<0.1$}
\end{flushleft}
\label{tab:zinb-cited}
\end{table}

\begin{table}[hb]
\centering
\scriptsize
\renewcommand{\arraystretch}{1.2}
\begin{tabular}{lcc}
\toprule
\textbf{Variable} & \textbf{Estimate (SE)} & \textbf{Signif.} \\
\midrule
\multicolumn{3}{l}{\textit{Main Model}} \\
Binoculars Score & $-1.790\ (0.307)$ & $^{***}$ \\
Post-GPT & $0.527\ (0.421)$ &  \\
Post-GPT × Binoculars Score & $0.346\ (0.419)$ &  \\
West & $1.175\ (0.476)$ & $^{**}$ \\
West × Binoculars Score & $-0.002\ (0.479)$ &  \\
Post-GPT × West & $-0.125\ (0.704)$ &  \\
Post-GPT × West × Binoculars & $-0.607\ (0.707)$ &  \\
h-index & $0.0022\ (0.0001)$ & $^{***}$ \\
SJR Q2 & $-0.672\ (0.025)$ & $^{***}$ \\
SJR Q3 & $-1.585\ (0.040)$ & $^{***}$ \\
SJR Q4 & $-2.307\ (0.045)$ & $^{***}$ \\
Exposure Score & $2.099\ (0.070)$ & $^{***}$ \\
\midrule
\multicolumn{3}{l}{\textit{Inflation Model}} \\
Binoculars Score & $1.513\ (0.511)$ & $^{***}$ \\
h-index & $-0.0031\ (0.0003)$ & $^{***}$ \\
Exposure Score & $-38.513\ (2.194)$ & $^{***}$ \\
SJR Q2 & $0.425\ (0.082)$ & $^{***}$ \\
SJR Q3 & $0.930\ (0.136)$ & $^{***}$ \\
SJR Q4 & $2.097\ (0.178)$ & $^{***}$ \\
\bottomrule
\end{tabular}
\caption{Zero-inflated negative binomial model predicting cross-hemispheric citation count.}
\vspace{-1em}
\begin{flushleft}
\footnotesize{$N = 97{,}995$, Log Likelihood = $-138{,}181$, Wald $\chi^2 = 11{,}838$, $p<0.001$. $^{***}p<0.001$, $^{**}p<0.01$, $^{*}p<0.1$}
\end{flushleft}
\label{tab:zinb-citediff}
\end{table}

\end{document}